\documentclass[12pt]{article}
\usepackage{graphicx,amssymb,amsfonts,amsmath,epsfig,color}

\newcounter{mnotecount}

\newcommand{\mnotex}[1]
{\protect{\stepcounter{mnotecount}}$^{\mbox{\footnotesize $\bullet$\themnotecount}}$
\marginpar{
\raggedright\tiny\em
$\!\!\!\!\!\!\,\bullet$\themnotecount: #1} }

\makeatletter
\DeclareSymbolFont{AMSb}{U}{msb}{m}{n}
\DeclareSymbolFontAlphabet{\mathbb}{AMSb}
\renewcommand{\section}{\@startsection{section}{1}{\z@}%
                                    {-7ex \@plus -1ex \@minus -.2ex}%
                                    {2.5ex \@plus.2ex}%
                                    {\normalfont\large\scshape\centering}}
\renewcommand{\subsection}{\@startsection{subsection}{2}{\z@}%
                                       {-5ex \@plus -1ex \@minus -.2ex}%
                                       {1.5ex \@plus.2ex}%
                                       {\normalfont\normalsize\scshape}}
\renewcommand{\subsubsection}{\@startsection{subsubsection}{3}{\z@}%
                                       {-5ex \@plus -1ex \@minus -.2ex}%
                                       {1.5ex \@plus.2ex}%
                                       {\normalfont\normalsize\scshape}}

\renewcommand\@seccntformat[1]{\ignorespaces\csname #1name\endcsname\space
                               \csname the#1\endcsname.\quad}   
\newdimen\captionmargin
\newdimen\captionindent
\newdimen\captionwidth
\newcommand{\captionfont}{\slshape}
\newcommand\@captionlabel[1]{\textsc{#1:}\space}
\long\def\@makecaption#1#2{%
  \vskip\abovecaptionskip
  \captionwidth\hsize
  \advance\captionwidth -2\captionmargin
  \sbox\@tempboxa{\@captionlabel{#1}\captionfont #2}%
  \ifdim \wd\@tempboxa >\captionwidth
    \ifdim\captionindent>\z@
      \advance\captionwidth -\captionindent
      \hskip\captionindent
    \fi
    \hskip\captionmargin
    \parbox[t]{\captionwidth}{\leavevmode\hskip-\captionindent
      \@captionlabel{#1}\captionfont #2}%
  \else
    \global \@minipagefalse
    \hb@xt@\hsize{\hfil\box\@tempboxa\hfil}%
  \fi
  \vskip\belowcaptionskip}
\def\eqnarray{%
   \stepcounter{equation}%
   \def\@currentlabel{\p@equation\theequation}%
   \global\@eqnswtrue
   \m@th
   \global\@eqcnt\z@
   \tabskip\@centering
   \let\\\@eqncr
   $$\everycr{}\halign to\displaywidth\bgroup
       \hskip\@centering$\displaystyle\tabskip\z@skip{##}$\@eqnsel
      &\global\@eqcnt\@ne$\;\hfil{##}$\hfil
      &\global\@eqcnt\tw@$\;\displaystyle{##}$\hfil\tabskip\@centering
      &\global\@eqcnt\thr@@ \hb@xt@\z@\bgroup\hss##\egroup
         \tabskip\z@skip
      \cr}
\ifcase \@ptsize
\or
\or
\fi
  \divide\@tempdima\baselineskip
  \@tempcnta=\@tempdima
\makeatother
\begin{document}

\renewcommand{\theequation}{\arabic{section}.\arabic{equation}}
\renewcommand{\thefigure}{\arabic{figure}}
\newcommand{\gapprox}{%
\mathrel{%
\setbox0=\hbox{$>$}\raise0.6ex\copy0\kern-\wd0\lower0.65ex\hbox{$\sim$}}}
\textwidth 165mm \textheight 220mm \topmargin 0pt \oddsidemargin 2mm
\def\ib{{\bar \imath}}
\def\jb{{\bar \jmath}}

\newcommand{\ft}[2]{{\textstyle\frac{#1}{#2}}}
\newcommand{\be}{\begin{equation}}
\newcommand{\ee}{\end{equation}}
\newcommand{\bea}{\begin{eqnarray}}
\newcommand{\eea}{\end{eqnarray}}
\newcommand{\Identity}{{1\!\rm l}}
\newcommand{\cx}{\overset{\circ}{x}_2}
\def\CN{$\mathcal{N}$}
\def\CH{$\mathcal{H}$}
\def\hg{\hat{g}}
\newcommand{\bref}[1]{(\ref{#1})}
\def\espai{\;\;\;\;\;\;}
\def\zespai{\;\;\;\;}
\def\avall{\vspace{0.5cm}}
\newtheorem{theorem}{Theorem}
\newtheorem{acknowledgement}{Acknowledgment}
\newtheorem{algorithm}{Algorithm}
\newtheorem{axiom}{Axiom}
\newtheorem{case}{Case}
\newtheorem{claim}{Claim}
\newtheorem{conclusion}{Conclusion}
\newtheorem{condition}{Condition}
\newtheorem{conjecture}{Conjecture}
\newtheorem{corollary}{Corollary}
\newtheorem{criterion}{Criterion}
\newtheorem{defi}{Definition}
\newtheorem{example}{Example}
\newtheorem{exercise}{Exercise}
\newtheorem{lemma}{Lemma}
\newtheorem{notation}{Notation}
\newtheorem{problem}{Problem}
\newtheorem{prop}{Proposition}
\newtheorem{rem}{{\it Remark}}
\newtheorem{solution}{Solution}
\newtheorem{summary}{Summary}
\numberwithin{equation}{section}
\newenvironment{pf}[1][Proof]{\noindent{\it {#1.}} }{\ \rule{0.5em}{0.5em}}
\newenvironment{ex}[1][Example]{\noindent{\it {#1.}}}

\thispagestyle{empty}


\begin{center}

{\LARGE\scshape An analogue first law for general closed marginally trapped surfaces}
\vskip10mm

\textsc{Ram\'{o}n Torres\footnote{E-mail: ramon.torres-herrera@upc.edu}}

\par\bigskip
{\em Dept. de F\'{i}sica Aplicada, Universitat Polit\`{e}cnica de Catalunya, Barcelona, Spain.}\\[.1cm]

\vspace{5mm}

\end{center}

\begin{abstract}
We formulate an analogue transverse first law for general closed marginally trapped surfaces in arbitrary spacetimes. The construction is intrinsically quasi-local and is attached directly to an individual marginally trapped surface, rather than to a preferred horizon worldtube. Taking the Hawking energy as the internal energy and an invariant effective surface gravity associated with the marginally trapped surface as the quantity controlling the thermal term, we derive a balance law in which the variation of energy splits into a generalized heat contribution and a total work contribution. 
In this way, the resulting law provides a codimension-two, transverse counterpart to existing horizon-based formulations of black-hole thermodynamics. We show that the formalism reproduces the expected results for round spheres in spherically symmetric spacetimes.
We then examine semiclassical equilibrium and evaporating regimes, and extend the analysis to non-spherically symmetric marginally trapped surfaces in Kerr. These examples indicate that the framework remains applicable in situations where a horizon-based treatment is either nonunique or technically cumbersome, and suggest that closed marginally trapped surfaces provide a natural arena for a genuinely quasi-local thermodynamics of black holes.
\end{abstract}




\setcounter{equation}{0}

\section{Introduction}

The thermodynamics of black holes emerged from the realization that the classical mechanics of stationary horizons in the framework of General Relativity is governed by laws that are formally identical to the ordinary laws of thermodynamics. In its standard form \cite{BardeenCarterHawking1973,Smarr1973}, the first law of black-hole mechanics relates neighboring stationary Kerr--Newman solutions through
\begin{equation}
\delta M=\frac{\kappa}{8\pi}\,\delta A+\Omega_H\,\delta J+\Phi_H\,\delta Q,
\end{equation}
where $M$, $J$ and $Q$ are the conserved charges measured at infinity, $A$ is the horizon area, $\kappa$ is the surface gravity, and $\Omega_H$ and $\Phi_H$ are the horizon angular velocity and electric potential, respectively. Bekenstein's proposal that horizon area should be interpreted as entropy, together with Hawking's discovery that black holes radiate thermally with temperature $T_H=\kappa/2\pi$, turned that formal analogy into genuine thermodynamics \cite{Bekenstein1973,Haw75}. Later developments, such as the Noether-charge formulation of black-hole entropy, clarified that the first law is deeply tied to quasi-local geometric data at the horizon even when its traditional derivation is performed within a stationary framework \cite{Wald1993,IyerWald1994}.

Historically, however, the first law was formulated for equilibrium or near-equilibrium situations and for horizons singled out by stationarity. This is conceptually restrictive. Astrophysical black holes form, accrete, merge, ring down and evaporate. In such regimes neither a global event horizon nor a Killing horizon is the most convenient object on which to base a local thermodynamic description. This observation motivated the search for quasi-local and dynamical generalizations of black-hole mechanics.

A major step in that direction was provided by Ashtekar and collaborators by introducing the \textit{isolated-horizon} framework \cite{AshtekarBeetleFairhurst1999,AshtekarFairhurstKrishnan2000} that replaces the global notion of a Killing horizon by a quasi-local null hypersurface that is itself in equilibrium, while still allowing radiation and dynamical processes in the exterior region. In that setting one obtains a local version of the zeroth and first laws without imposing global stationarity. Later, they proposed the complementary \textit{dynamical-horizon} framework \cite{AshteKris,AshtekarKrishnan2004} which extends the discussion to spacelike hypersurfaces foliated by marginally trapped surfaces, yielding flux laws for energy and angular momentum and first-law-type balance relations for growing black holes. More recently, this quasi-local-horizon program has been significantly broadened: the modern notion of dynamical horizon segments accommodates more general signatures and emphasizes the direct relation between horizon-area change and local fluxes, as well as the role of quasi-local horizons in binary mergers, approach to equilibrium, and evaporation \cite{AshtekarKrishnan2025}. These ideas have had a decisive impact on mathematical relativity, numerical relativity and quantum-gravity approaches to black-hole entropy.

In parallel, Hayward developed a different but equally influential program based on \textit{trapping horizons}. Starting from his general laws of black-hole dynamics and his unified first law in spherical symmetry \cite{Hayward1994,Hkappa}, and culminating in his formulation of energy and entropy conservation for dynamical black holes \cite{Hay2004}, he obtained first-law-type relations in which matter and gravitational-radiation fluxes appear on the same footing. In Hayward's framework, the horizon is described quasi-locally as a hypersurface foliated by marginal surfaces, and the corresponding balance laws are written directly in terms of geometric quantities intrinsic and normal to those foliating surfaces. These results constitute one of the clearest generalizations of black-hole thermodynamics away from strict equilibrium.

Despite their importance, these generalizations leave open a fundamental methodological issue that is directly relevant to the present work. Historically, quasi-local frameworks were tied to specific classes of hypersurfaces: isolated horizons were restricted to null, intrinsically equilibrium objects, while original dynamical horizons were spacelike by definition \cite{AshteKris,AshtekarKrishnan2004}. Furthermore, trapping-horizon constructions presupposed a preferred horizon hypersurface and, in practice, often required an adapted auxiliary structure \cite{Hay2004}. Recent advances \cite{AshtekarKrishnan2025,AshtekarParaizoShu2025} have elegantly broadened this paradigm by introducing a more flexible segmentation language. A quasi-local horizon may now contain null equilibrium segments continuously joined to non-null dynamical segments that are either spacelike (e.g., during accretion) or timelike (e.g., during evaporation). 

Consequently, the limitation addressed in the present paper is not that far-from-equilibrium processes like evaporation lie outside the modern quasi-local-horizon paradigm. Rather, the central issue is that these frameworks---both classical and modern---are primarily formulated as evolution laws \emph{along} a chosen three-dimensional horizon worldtube. By contrast, our aim is to formulate a first-law-type balance relation directly on an individual \textit{closed marginally trapped surface} (codimension-two), prior to and independently of the choice of any preferred encompassing null, spacelike, or timelike tube. 

This conceptual shift from a longitudinal tube to a transversal surface also resolves significant practical difficulties. For instance, explicit implementation of worldtube-based formalisms (especially Hayward's) beyond highly symmetric situations can be technically demanding; in particular, for the exact Kerr spacetime, no globally useful dual-null foliation is known in closed form. Furthermore, in semiclassical evaporation, the longitudinal fluxes encoded in the Unruh state are notoriously delicate near the horizon in static frames \cite{VisserU}, which heavily complicates any thermodynamic formulation based directly on longitudinal horizon evolution. These remarks do not diminish the immense scope of the existing frameworks; rather, they clearly indicate the need for a complementary, purely surface-based transversal viewpoint.

As previously stated, the natural geometric arena for such a transversal viewpoint is furnished by closed marginally trapped surfaces. Geometrically, a closed marginally trapped surface (MTS) is a closed, spacelike codimension-two surface for which one future-directed null expansion vanishes while the other is non-positive. In black-hole physics, these surfaces serve as the fundamental building blocks that foliate apparent horizons, trapping horizons, marginally trapped tubes, and dynamical horizons. They are therefore conceptually and mathematically more primitive than any particular three-dimensional horizon worldtube constructed from them. From a strictly quasi-local perspective, it is the individual MTS itself---rather than its extended temporal history along a worldtube---that encodes the irreducible geometric data required to define area, entropy, quasi-local energy concentration, and surface gravity.

Moreover, the geometry of MTSs is far richer than the textbook picture of a single preferred black-hole boundary might suggest. The literature initiated and developed by Senovilla and collaborators shows that trapped and marginally trapped structures are highly non-unique, that many marginally trapped tubes can pass through a given marginally trapped surface, and that in spherical symmetry closed trapped surfaces can extend across both sides of the usual apparent 3-horizon associated with black holes and even penetrate regions whose past is flat \cite{Senovilla2011,B&S0,B&S,Senovilla2014Remarks}. In this sense, marginally trapped structures are not confined to one distinguished hypersurface: they proliferate around the black-hole region. This is precisely the regime in which a thermodynamical description based on a specific horizon type becomes too narrow.

The starting point of the present paper is the proposal in \cite{SenovillaTorres2015} for an invariant notion of surface gravity that can be assigned to a compact MTS once a suitable family of surrounding surfaces is chosen. It reproduces the usual Killing-horizon value in Schwarzschild, reduces to Hayward's Kodama-based expression in spherical symmetry, and remains meaningful in genuinely non-symmetric situations \cite{SenovillaTorres2015}. The central idea of this work is to take that result as the seed of a broader thermodynamical structure.

Our proposal is to formulate an analogue first law directly on a general \emph{closed} MTS, using Hawking energy as the natural quasi-local internal energy and the effective surface gravity of \cite{SenovillaTorres2015} as the quantity controlling the thermal term. The resulting law is not an equilibrium phase-space law between neighboring stationary solutions, nor an evolution law tied to a preferred horizon hypersurface. Rather, it is a quasi-local \emph{transverse} balance law associated with deformations of the MTS along a chosen null direction. In this framework, the heat and work contributions are read from purely geometric and quasi-local quantities.
In this sense, the present construction is best viewed as complementary to the modern horizon-segment laws: instead of providing another balance law along a quasi-local horizon, it isolates an intrinsically codimension-two thermodynamical relation attached to the MTS itself.
This viewpoint is designed to apply uniformly to general MTSs, including situations where no Killing field exists, where the relevant horizon tube is not unique, and where the longitudinal evolution is obscured by dynamical or semiclassical complications.

The goals of this article are therefore the following. First, we derive an analogue first law for arbitrary closed marginally trapped surfaces, written entirely in terms of quasi-local geometric data. Second, we identify the corresponding notions of generalized heat and generalized work, separating the contribution associated with the effective surface gravity from the matter and gravitational work densities. Third, we show that the formalism reproduces the expected behavior in standard symmetric situations while remaining applicable beyond spherical symmetry. Fourth, we clarify how this intrinsically surface-based law relates to existing quasi-local-horizon thermodynamics, including recent far-from-equilibrium developments. Fifth, we illustrate that the same framework continues to be meaningful in semiclassical equilibrium and, crucially, in evaporating situations, where previous horizon-based formulations ---although now considerably more powerful than in the original isolated/dynamical-horizon setting--- are not always the most economical tools in practice. In this sense, the paper aims to provide a genuinely surface-based complement to the existing horizon-based generalizations of black-hole thermodynamics.

This paper is organized as follows. In Sec.\ref{GP} we review the geometrical preliminaries for spacelike two-surfaces in spacetime and fix the notation for null expansions, mean curvature and dual expansion vector. In Sec.\ref{secPP} we introduce the physical ingredients needed for the thermodynamical interpretation, namely Hawking energy, the energy concentration, the effective surface gravity and the associated work densities. In Sec.\ref{secFL} we derive the analogue first law for closed MTSs and interpret its various terms. In Sec.\ref{secSph} we study the case of round spheres in spherically symmetric spacetimes, with separate discussions of the equilibrium and evaporating semiclassical regimes. In Sec.\ref{secKerr} we analyze a non-spherically symmetric example, namely marginally trapped surfaces in Kerr, again considering both equilibrium and evaporating situations. Finally, Sec.\ref{secConc} contains the conclusions and a discussion of the scope and possible applications of the formalism.

\section{Geometrical Preliminaries}\label{GP}

Let $(\mathcal V, g)$ be a 4-dimensional causally orientable spacetime with metric signature $\{-,+,+,+\}$ and with local coordinates $\{x^\alpha\}$.
Let $\mathcal S$ denote a connected 2-dimensional surface with local intrinsic coordinates $\{\lambda^A\}$
imbedded in $\mathcal V$ by the $C^3$ parametric equations
\[
x^\alpha=\Phi^\alpha(\lambda^A).
\]
The tangent vectors $\vec{e}_A$ of $\mathcal S$  are locally given by
\[
\vec{e}_A\equiv e^\mu_A \frac{\partial}{\partial x^\mu}\rfloor_{\mathcal S} \equiv \frac{\partial \Phi^\mu}{\partial \lambda^A} \frac{\partial}{\partial x^\mu}\rfloor_{\mathcal S}
\]
so that the first fundamental form of $\mathcal S$ in $\mathcal V$ is
\[
\gamma_{AB}\equiv g_{\mu\nu}\rfloor_{\mathcal S} \frac{\partial \Phi^\mu}{\partial \lambda^A}\frac{\partial \Phi^\nu}{\partial \lambda^B}.
\]
We are interested in spacelike surfaces $\mathcal S$ in which case $\gamma_{AB}$ is positive definite. The two linearly independent null future-directed one-forms normal to $\mathcal S$ are denoted by $l^\pm_\mu$. They satisfy
\[
l^\pm_\mu e^\mu_A=0,\ \ \ l^+_\mu l^{+\mu}=0, \ \ \ l^-_\mu l^{-\mu}=0
\]
and, without loss of generality, we choose them to satisfy the convenient normalization condition
\begin{equation}\label{norm}
l^+_\mu l^{-\mu}=-1.
\end{equation}
The covariant derivatives on $(\mathcal V, g)$ and on $(\mathcal{S}, \gamma)$ are related through \cite{Kriele}\cite{ONeill}
\[
e^{\rho}_{A}\nabla_{\rho}e^{\mu}_{B}=\overline{\Gamma}^C_{AB} e^{\mu}_{C}-K^{\mu}_{AB}
\]
where $\overline{\Gamma}^C_{AB}$ are the coefficients of the Levi-Civita connection $\overline{\nabla}$ of $\gamma$ (so that $\overline{\nabla}\gamma=0$) and $K^{\mu}_{AB}$ is the shape tensor (also called second fundamental form vector or extrinsic curvature vector) of $\mathcal S$  in $(\mathcal V, g)$. The shape tensor is normal to ${\cal S}$ and thus it can be decomposed as
\[
K^{\mu}_{AB}=-K^-_{AB} l^{+\mu} -K^+_{AB} l^{-\mu},
\]
where $K^\pm_{AB}$ are called the two null (future) second fundamental forms of $\mathcal S$  in  $(\mathcal V, g)$
given by
\[
K^\pm_{AB}\equiv e^\nu_A e^\mu_B \nabla_\nu l^\pm_\mu , \hspace{2cm} K^\pm_{AB}=K^\pm_{BA} \, .
\]
The mean curvature vector of $\mathcal S$  in $(\mathcal V, g)$ \cite{Kriele}\cite{ONeill}
is the trace of the shape tensor
\[
H^{\mu}\equiv \gamma^{AB} K^{\mu}_{AB}
\]
where $\gamma^{AB}$ is the contravariant metric on $\mathcal S$  ($\gamma^{AC} \gamma_{CB}=\delta^A_B$).
Clearly, the mean curvature vector is orthogonal to $\mathcal S$ and can be written in the form
\[
H^{\mu}= -\theta^- l^{+\mu} -\theta^+ l^{-\mu},
\]
where
\[
\theta^\pm\equiv \gamma^{AB} K^{\pm}_{AB}
\]
are the traces of the null second fundamental forms, also called the (future) null expansions.
We will be specially interested in surfaces in which $\vec{H}\ (:\neq \vec{0})$ is null everywhere on $\mathcal S$, keeping its causal orientation (future or past) and pointing consistently along one of two null directions $l^\pm$. These surfaces are called marginally (future or past) trapped surfaces (MTS) and satisfy either $\{\theta^+=0, \theta^- \leq 0\}$ or $\{\theta^-=0, \theta^+ \leq 0\}$ for the future case (reverse inequalities for the past case). For concreteness, and unless stated otherwise, from now on we will tacitly assume that we are dealing with the first case when considering a MTS. We will also assume that the non-vanishing expansion is strictly negative for simplicity and to avoid unnecessary complications.

The {\em unique} vector field dual to $\vec{H}$ in the plane orthogonal to $\mathcal S$, called the \textit{dual expansion vector} \cite{Tung}\cite{SenoProc}, takes the form
\begin{equation}
\ast H^{\mu}= -\theta^- l^{+\mu} +\theta^+ l^{-\mu}.
\end{equation}
This vector field defines the (generically unique) direction with vanishing expansion of $\mathcal S$ \cite{Tung}\cite{SenoProc}.
$\ast \vec H$ is timelike for untrapped surfaces, spacelike for trapped surfaces and null (equal to $\vec H$) for MTS.
Moreover, in the framework of quasi-local Hamiltonians it defines the direction of a Hamiltonian flow at $\mathcal S$ \cite{Tung}\cite{Anco}. 
In particular, in spherically symmetric spacetimes and when $\mathcal S$ is chosen to be a round sphere, $\ast \vec{H}$ is parallel to the \textit{Kodama vector} \cite{kodama}. For our purposes, it is enough to note that, for untrapped surfaces,
\begin{equation}\label{GKO}
\hat u\ \equiv \frac{\ast\vec{H}}{\sqrt{- g(\ast\vec{H},\ast\vec{H})}}
\end{equation}
defines on $\mathcal S$ the 4-velocity of privileged observers with respect to $\mathcal S$, in the sense that they measure no expansion of ${\cal S}$.

\section{Physical Preliminaries}\label{secPP}

Given any compact MTS ${\cal S}$ without boundary, consider a local foliation $\mathcal N (t)$ of the spacetime by pieces of spacelike hypersurfaces in such a way that, for a given value $t_{0}$ of $t$, our MTS ${\cal S} \subset {\cal N}(t_{0})$. Construct then a local tube foliated by compact spacelike surfaces $\mathcal S(t)$, each of them lying in one of the $\mathcal N (t)$ ($\mathcal S(t_0)\ \equiv \mathcal S$).
\begin{figure}
\centering
\includegraphics[scale=.9]{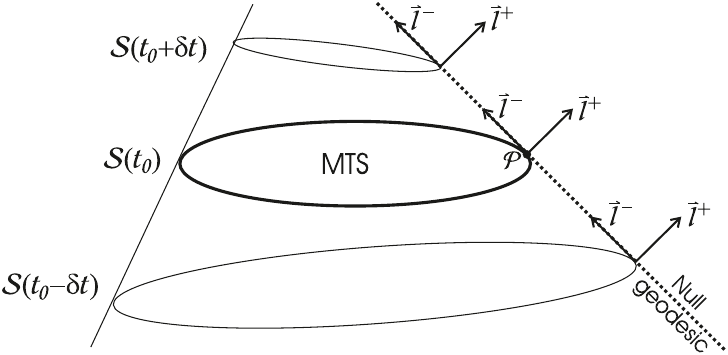}
\caption{\label{NS} A schematic representation showing some surfaces from the tube. The
lightlike geodesic with tangent vector $\vec l^-$ is described by a dashed line. For the
purposes of  this figure we are considering the case of compact surfaces $\mathcal S (t)$.}
\end{figure}

The tube is further constrained so that it contains the null geodesic generated at ${\cal P}$ by the tangent vector $\vec{l}^-$ normal to $\mathcal S(t_0)$ (see figure \ref{NS}), and such that each ${\cal S}(t)$ has a negative (respectively positive) expansion $\theta^{+}$ at their intersection with the null geodesic for $t>t_{0}$ (resp.\ for $t< t_{0}$). This is related to the outermost stability (or instability) of ${\cal S}$ \cite{AMS,AMS1}.

\textit{Hawking energy} on an arbitrary closed compact surface ${\cal S}$ is given by \cite{Hawking1968}
\begin{equation}\label{HE}
E=\frac{R}{16 \pi} \oint_{\mathcal S} \ast \left(^{(2)}\mathcal R + \frac{g(\ast H,\ast H)}{2}\right),
\end{equation}
where $^{(2)}\mathcal R$ is the curvature of $\mathcal S$ and 
\[
R =\sqrt{ A/(4 \pi)}
\]
is its \textit{areal radius}.
Note that $E$ is a genuinely quasi-local functional defined for any closed spacelike 2-surface, constructed solely from intrinsic geometry and the null expansions (hence boost-gauge invariant). It arises naturally as the Hamiltonian associated with the dual-expansion flow used to define our preferred observers, and in spherical symmetry it reduces to the Misner–Sharp mass.

The magnitude
\begin{equation}\label{defPsi}
\Psi\equiv \frac{E}{R}= \frac{1}{16 \pi} \oint_{\mathcal S} \ast \left(^{(2)}\mathcal R +\frac{g(\ast H,\ast H)}{2}\right)
\end{equation}
will be called the \textit{energy concentration} since it provides us with a measure of how concentrated is Hawking energy inside the surface\footnote{For this reason it is sometimes called the \textit{compactness}, a ratio of central importance in the geometric flow literature\cite{SamuelRoyChowdhury2008}. We avoid that name here to prevent confusion with the topological mathematical terminology.}.

In a future MTS $\theta_{+} = 0$ and $g(*H, *H) = 0$. Then, on the MTS $\Psi$ reduces exactly to the topological piece:
\begin{equation}\label{HEMTS}
\Psi\mid_{MTS} = \frac{1}{16\pi} \oint_{MTS} * {}^{(2)}\mathcal R = \frac{\chi(MTS)}{4} = \frac{1 - g}{2},
\end{equation}
where $\chi(MTS)$ is the Euler characteristic of the MTS
and $g$ is the genus.

The \textit{effective surface gravity} \cite{SenovillaTorres2015} associated with a point on a compact (future)
MTS included in the family of chosen surfaces $\mathcal S (t)$ can be written as
\begin{equation}\label{ESG}
\kappa\equiv -\sqrt{\frac{A(\mathcal S_{MTS})}{16\pi}}\  l^{-\alpha}\nabla_\alpha \theta^+\rfloor_{MTS}
\end{equation}

Important properties of this quantity are that it is geometrically invariant as well as independent of the parametrization of $\vec l^\pm$.

A surface gravity for the whole MTS can be defined as the \textit{mean effective surface gravity} $\overline{\kappa}$ on the surface, i.e.,
\begin{equation}\label{mesg}
\overline{\kappa}\equiv\frac{1}{A} \oint_{\mathcal S} \ast\ \kappa.
\end{equation}

The existence of an effective surface gravity $\kappa$ suggests the existence of an effective local \textit{Temperature} $T$ for the MTS. However, care must be taken in the following comments since it is only appropriate to talk about a temperature if the MTS is part of a marginally trapped tube with slowly varying effective surface gravity, see in this respect \cite{B,B1,Barc}. Only in this case one could define an \textit{effective temperature} $T$ on a point of the compact MTS and a \textit{mean effective temperature} $\overline{T}$ for the whole MTS as
\begin{equation}\label{defT}
T\equiv\frac{\kappa}{2 \pi}\ \ \mbox{and} \ \ \overline{T}\equiv\frac{\overline\kappa}{2 \pi},
\end{equation}
respectively.

\textit{Remark}: Note that the effective temperature $T$ is intrinsically tied to the chosen foliation. Physically, this reflects the observer-dependence of temperature. To extract a unique scalar field $T$, the boost freedom is fixed by our chosen foliation of the spacetime, which corresponds physically to the frame of the macroscopic privileged observers evaluating the thermodynamic system.

Provided an effective energy momentum tensor $T_{\mu\nu}=G_{\mu\nu}/(8 \pi)$, one can define the \textit{matter work density} $\omega_m$ (work per unit volume) on a point of $\mathcal S$ as \cite{Hay2004}
\[
\omega_m = T_{\mu \nu} l^{+\mu} l^{-\nu}.
\]
Let us define  \cite{Hay2004}\cite{AshteKris}
\begin{equation}\label{zeta}
\zeta^\alpha=-\gamma^{\alpha\beta} l^{-\nu} \nabla_\nu l^+_\beta
\end{equation}
with
\[
\gamma^{\alpha\beta}=\gamma^{AB} \frac{\partial \Phi^\alpha}{\partial \lambda^A}\frac{\partial \Phi^\beta}{\partial \lambda^B},
\]
where $\gamma^{AB}$ is the inverse of the first fundamental form of $\mathcal{S}$.
The vector $\zeta$ is a spacelike 4-vector tangent to $\mathcal S$ that measures how the outgoing null normal $\vec{l}^+$ "tilts" or "drifts" tangentially to the surface as one moves transversally off the surface along the ingoing null ray $\vec{l}^-$.

The \textit{gravitational work density} $\omega_g$ on a point of $\mathcal S$ is \cite{Hay2004}
\begin{equation}\label{omg}
\omega_g =\frac{|\zeta|^2}{8 \pi}.
\end{equation}
Since $|\zeta|^2=\zeta^\alpha\ \zeta_\alpha\geq 0$ the magnitude $\omega_g$ has the two desirable properties of being non-negative and (from (\ref{zeta})) of being independent of the parametrization of the normal null vectors satisfying $\vec l^+ \cdot \vec l^-=-1$.

The (total) \textit{work density} on a point of $\mathcal S$ is
\[
\omega \equiv \omega_m + \omega_g.
\]
As previously, one can define the corresponding \textit{mean work densities} ($\overline\omega_m$,  $\overline\omega_g$ and  $\overline\omega$) for the whole surface by using
\begin{equation}\label{mean}
\overline\omega_{()}=\frac{1}{A} \oint_{\mathcal S} \ast\ \omega_{()}.
\end{equation}

\section{The analogue transverse first law for closed MTSs}\label{secFL}

In classical thermodynamics the first law relates the changes in internal energy of the system with the changes in heat supply and work (`$\delta E=\delta Q+\delta W$'). Our goal here is to write an analogue law for a MTS. The expression for the law should depend on which MTS we are evaluating it, but it should be independent of any specific point of the MTS. In order to do this we will consider Hawking energy as the internal energy in the MTS and we will use the \emph{mean} surface gravity in the combination $\overline\kappa\ \delta A/(8 \pi)$ as a contribution to the heat term (in analogy with previous expressions for BH thermodynamics). We start by considering variations evaluated on a MTS along the chosen family of surfaces: $\delta\equiv l^{-\alpha}\nabla_\alpha$. We have $\delta E = \delta(R\Psi) = \Psi \delta R + R \delta\Psi$, $\delta A = 8 \pi R \delta R$ and  $\delta V = A \delta R$, where $V=(4/3) \pi R^3$ is the \textit{areal volume}. This leads to
\begin{equation}\label{dE}
\delta E-\frac{\overline{\kappa}}{8 \pi} \delta A=\frac{1}{A}\left(\frac{E}{R}-\overline{\kappa} R \right) \delta V +R \delta \Psi.
\end{equation}
The quantity in parenthesis can be written on the MTS with the help of (\ref{HEMTS}), (\ref{ESG}) and (\ref{mesg}) as
\[
\mathcal{W} \equiv \frac{E}{R}-\overline{\kappa} R = \frac{1}{16 \pi} \oint_{\mathcal S} \ast\ (^{(2)}\mathcal{R}+
2 l^{-\alpha} \nabla_\alpha \theta^+).
\]
On the other hand, using the definition of the effective energy-momentum tensor, one finds that on the MTS \cite{Hay2004}
\[
l^{-\alpha} \nabla_\alpha \theta^+=- ^{(2)}\mathcal{R}/2 + |\zeta|^2-D_i \zeta^i +8\pi T_{\mu\nu} l^{+\mu} l^{-\nu},
\]
where $D$ is the covariant derivative corresponding to $\gamma$. Using this expression $\mathcal{W}$ can be rewritten as
\[
\mathcal{W}=\frac{1}{8\pi} \oint_{\mathcal S} \ast\ |\zeta|^2 + \oint_{\mathcal S} \ast\ T_{\mu\nu} l^{+\mu} l^{-\nu}=A (\overline{\omega}_g +\overline{\omega}_m)=A \overline{\omega},
\]
where the Gauss divergence theorem has been used on the closed MTS to eliminate the $D_i \zeta^i$ term.
In this way, we see that $\mathcal{W}/A$ is the mean work density on the MTS.

The relationship (\ref{dE}) 
can now be written as
\begin{equation}
\delta E\stackrel{MTS}{=}\frac{\overline{\kappa}}{8 \pi} \delta A+ R \delta \Psi+\overline{\omega} \delta V 
\end{equation}

Clearly the term controlled by a work density and proportional to a volumetric displacement is the \textit{variation of total work}
\[
\boxed{\delta W_- \equiv \overline{\omega} \delta V}.
\]
The part of the change in Hawking energy that does not come from the variation of the total work is the \textit{generalized transversal heat exchange} 
\begin{equation}\label{dQ-}
\delta Q_-\equiv \frac{\overline{\kappa}}{8 \pi} \delta A+ R \delta \Psi.
\end{equation}
This is natural. This decomposition shows that a transverse deformation of the MTS generally changes both its area and the energy concentration. On the one hand, the term $(\overline{\kappa}/8\pi)\,\delta A$ is the usual purely entropic thermal agitation or \textit{bare heat}. On the other hand,
$\Psi$ characterizes the energy concentration of the quasi-local energy enclosed by a surface of a given areal radius $R$. Consequently, the term $R\delta\Psi$ represents a geometric analogue of latent heat —specifically, the configurational energy cost associated with deformations of the boundary surface. This indicates that, at a quasi-local level, the gravitational degrees of freedom encoded in the curvature actively participate as an effective thermodynamic reservoir.

It is instructive to analyze the physical content of $\delta Q_{-}$ in greater detail.
From (\ref{defPsi}) and (\ref{ESG}) one has 
\[
\delta\Psi\Big|_{MTS}=
-\frac{1}{8\pi R}\int_{MTS} {}^\ast\big(\kappa\,\theta_-\big)\; .
\]

Using now this result and (\ref{dQ-}) one gets the useful identity
\begin{equation}\label{dQ-k}
\delta Q_- = \frac{1}{8\pi}\int_{MTS}{}^\ast\Big[\theta_-(\bar\kappa-\kappa)\Big].
\end{equation}
Note that the expansion scalar $\theta_-$ dictates the fractional rate of change of the local area element, meaning $\delta(dA) = \theta_- dA$.
Thus,
if the MTS is uniformly symmetric, the local gravitational pull is identical everywhere across the surface and the integrand vanishes identically, leading to $\delta Q_- = 0$. However, when the MTS is distorted (e.g., during a ringdown phase, absorbing matter asymmetrically, or rotating), so that $\kappa$ varies across the surface, the integrand need not vanish and $\delta Q_-$ can be nonzero.
In that case, $\delta Q_-$ measures the contribution associated with spatial inhomogeneity of the effective surface gravity, weighted by the local area variation.

On the other hand, if one defines the entropy of a MTS with the usual Bekenstein-Hawking form $S = A/4$, the term $\frac{1}{4}\theta_- dA$ is physically equivalent to the local geometric variation in entropy, $\delta s\equiv \delta (dS)$. Using now the proportionality between $\kappa$ and $T$ (\ref{defT}) in equation (\ref{dQ-k}), it reduces exactly to:
\[
\boxed{ \delta Q_- = \int_{MTS} (\overline{T} - T) \, \delta s}.
\]
This structure appears ubiquitously in macroscopic non-equilibrium thermodynamics. It mathematically reveals that the MTS acts as a spatially extended, non-isothermal thermodynamic medium. 
In this rewriting, $(\overline{T} - T)$
denotes the deviation of the local effective temperature from its surface average, and the integral weights that deviation by the local entropy variation $\delta s$. Accordingly, $\delta Q_-$
may be read as an entropy-weighted measure of transverse temperature inhomogeneity on the MTS.

With these definitions, (\ref{dE}) takes the form of an \textit{analogue transverse first law}:
\begin{equation}\label{firstlaw}
\boxed{\delta E\stackrel{MTS}{=} \delta Q_- + \delta W_-}.
\end{equation}

\textit{Remark}: This equation should be understood as a deformation identity for the Hawking energy along the ingoing-null deformation $\delta$ associated with the chosen surrounding family of surfaces around the MTS.

\section{Example: Round Spheres in Spherically Symmetric space-times}\label{secSph}

The general spherically symmetric metric in advanced Eddington-Finkelstein coordinates takes the form
\begin{equation}\label{SEEF}
ds^2=-e^{2 \beta} \left( 1-\frac{2 m}{r}\right) du^2+2 e^{\beta} du dr+ r^2 (d\vartheta^2+\sin^2\vartheta d\varphi^2),
\end{equation}
where $\beta=\beta(u,r)$ and $m=m(u,r)$.
We choose to work with the family of round spheres (defined by constant values of $u$ and $r$). The areal radius in this case is simply $R=\sqrt{A/(4 \pi)}=r$. The null vector fields normal to the round spheres, i.e., the radial null vectors can be written as
\begin{equation}\label{ltd}
\vec l^+=\frac{\partial}{\partial u}+ \frac{e^{\beta}}{2} \left( 1-\frac{2 m}{r}\right) \frac{\partial}{\partial r}\ \ \ \mbox{and}\ \ \ \vec l^-=- e^{-\beta} \frac{\partial}{\partial r}
\end{equation}
The expansion corresponding to the outgoing null geodesics takes the expression
\[
\theta^+= \frac{e^{ \beta}}{r} \left( 1-\frac{2 m}{r}\right).
\]
confirming that, as is well-known, a round sphere in which $r=2 m$ is a MTS so that the hypersurface $r=2 m$ defines an apparent 3-horizon (A3H).
The dual expansion vector in this case (parallel to the Kodama vector) is
\[
*H=\frac{2e^{-\beta}}{r}\,\partial_u.
\]
Thus, the 4-velocity of the privileged observers is
\begin{equation}\label{usphe}
\hat u=\frac{e^{-\beta}}{\sqrt{1-2m/r}}\,\partial_u.
\end{equation}

For a round sphere of radius $r$,
\[ 
{}^{(2)}R = \frac{2}{r^2},\  \theta_{+}\theta_{-} = -\frac{2}{r^2} \left( 1 - \frac{2m}{r} \right)\ \text{and}\ \Psi = \frac{m(u, r)}{r},
\]
and consequently the Hawking energy $E = R\Psi = m$, i.e. Misner–Sharp mass, as expected.

We obtain the effective surface gravity along the A3H as
\begin{equation}\label{kSph}
\kappa= -\sqrt{\frac{A(\mathcal S_{A3H})}{16\pi}}\  l^{-\alpha}\nabla_\alpha \theta^+\rfloor_{A3H}=\left.\frac{1-2 m' }{4 m}\right\rfloor_{A3H},
\end{equation}
where $m'$ stands for the partial derivative of $m$ with respect to $r$. Due to spherical symmetry one has $\overline\kappa =\kappa$.
The local \textit{effective temperature} (when applicable) will also coincide with the \textit{mean effective temperature} and will take the form
\begin{equation}\label{TSph}
T=\left.\frac{1-2 m' }{8 \pi m}\right\rfloor_{A3H}.
\end{equation}

For round spheres in a spherically symmetric spacetime, the relevant "rotation/twist" information of the null normals vanishes, i.e., $\zeta = 0 \Rightarrow \omega_g = 0.$

\subsection{Equilibrium case}\label{subsecEquil}

To test the robustness of our analogue first law (\ref{firstlaw}) in a semiclassical regime, let us consider a spherically symmetric geometry that reduces to a Schwarzschild black hole at leading order and which is in strict thermal equilibrium. Following York \cite{York}, this requires enclosing the black hole within a reflecting cavity of radius $r_{box}$, maintained at the local Tolman temperature
\[
T_w(r_{\rm box})=\frac{T_{KV}}{\sqrt{-g_{uu}(r_{\rm box})}},
\]
where $T_{KV}=\kappa_{KV}/(2\pi)$ is the Killing temperature associated with $\xi=\partial_u$ that satisfies 
$T_{KV} = e^{\beta(r_{A3H})} T$, with $T$ given by (\ref{TSph}). Then, the local Tolman temperature can be written as
\[
T_w = T\,e^{\beta(r_{A3H}) -\beta(r_{\rm box})}
\left(1-\frac{2m(r_{\rm box})}{r_{\rm box}}\right)^{-1/2}.
\]

In this way, the heat capacity of the total system becomes positive and the canonical ensemble becomes strictly stable.
The quantum field reaches equilibrium in a Hartle-Hawking Vacuum State ($|H\rangle$). This state describes a plasma of Hawking radiation and vacuum polarization in perfect equilibrium with the horizon.
Thus, the classical matter $T_{\mu\nu} = 0$ is replaced by the renormalized tensor $\langle T_{\mu\nu} \rangle_H$ and the semiclassical Einstein equation is $G_{\mu\nu}=8\pi \langle T_{\mu\nu} \rangle_H$. In the equilibrium situation described, it has components \cite{Page82,VisserH}
\[
\langle T_{\hat\mu\hat\nu} \rangle_H =\operatorname{diag}(\rho, p_r, p_\perp,p_\perp)
\]
in a local Lorentz basis attached to the fiducial static observers (FIDOs), who coincide with the privileged observers with 4-velocity (\ref{usphe}) in this equilibrium case. 

Just as an illustration, note that, in a fixed-background Schwarzschild {\emph approximation} and when considering the action of conformally coupled massless scalars, these quantities are given by the Page–Visser result \cite{Page82,VisserH}. Thus, for instance, 
\[
\rho\simeq -36 p_\infty, \ \   \text{  where  }\ \ p_{\infty} \equiv \frac{\hbar}{90(16\pi)^2(2 m)^4}
\]
is the blackbody radiation pressure at the Hawking temperature measured at infinity.

However, in a fully self-consistent semiclassical treatment, it should instead be determined from the corresponding static backreacted solution.

According to the foundational works \cite{Page82,C&F77,Cande80}, for this quantum tensor to be non-divergent and regular exactly at the event horizon,
the quantum energy density evaluated at the horizon ($\rho$) and the radial pressure ($p_r$) must satisfy a strict kinematic constraint in the limit as the MTS is approached:
\[
p_r = -\rho.
\]
Using this result and (\ref{norm}) the matter work density of a MTS on the horizon will simply be:
$$
\omega_m = \langle T_{\mu\nu}\rangle_H\, l^{+\mu} l^{-\nu}=\frac{1}{2}(\rho-p_r)=\rho.
$$

Thus, in effect, the term $\omega_m$ extracts exactly the energy density of Hawking radiation/vacuum polarization evaluated on the MTS. On the other hand, the variation of work along the ingoing-null deformation $\delta W_- = \overline{\omega} \delta V=\rho \delta V$ can be recast in the suggestive thermodynamic form
\begin{equation}\label{Wsphe}
\delta W_-=-p_r \delta V.
\end{equation}

From 
(\ref{dQ-k}) one gets the expected result for a uniform MTS: 
\begin{equation}\label{Qsphe}
\delta Q_- = 0,
\end{equation}
which demonstrates that traversing these MTSs is a strictly adiabatic geometric process.

Using (\ref{Wsphe}) and (\ref{Qsphe}) the transverse first law (\ref{firstlaw}) in a self-consistent backreacted equilibrium geometry can be written for this case as 
\begin{equation} \label{dEHH}
\delta E = \rho \delta V.
\end{equation} 
This is a natural result: Because the spacetime in this situation is filled with a thermal plasma, the quasi-local Hawking energy $E$ depends on the radius. Stepping transversely inwards across the MTS reduces the enclosed volume ($\delta V < 0$), 
leaving a thin shell of quantum gas outside our control surface. 
Moreover, since $\rho<0$ one has $\delta E>0$, i.e., the quasi-local energy increases under inward steps because one is “leaving behind” negative energy density near the MTS. We see that our quasi-local framework remains fundamentally exact even when accommodating the semiclassical backreaction of Hawking radiation.

\subsection{Evaporating case}

Let us now analyze the robustness of our transverse first law during active black hole evaporation. A physical black hole formed from collapse is described by a dynamically backreacting geometry and a quantum ``in-vacuum'' state ($|in\rangle$). 

In a local Lorentz basis attached to the privileged observers, the symmetries of this case dictate that the renormalized stress-energy tensor in this model takes the form:
\begin{equation}
\langle T_{\hat{\mu}\hat{\nu}}\rangle\equiv
\begin{pmatrix}
\rho & f & 0 & 0\\
f & p_{r} & 0 & 0\\
0 & 0 & p_{\perp} & 0\\
0 & 0 & 0 & p_{\perp}
\end{pmatrix}.
\end{equation}

By the fundamental axioms of semiclassical gravity, any physical state yielding a well-defined backreaction must satisfy the local Hadamard condition \cite{DMP,WaldQFT}, even on a fully dynamical background. Because the state is Hadamard, covariant point-splitting mathematically guarantees that local boost-invariant quantities such as the transverse angular pressure ($p_{\perp}$)---which is invariant under radial boosts---and the scalar trace of the renormalized stress tensor ($\langle{T^{\mu}}_{\mu}\rangle$) derived from covariant point-splitting \cite{Chris76} are expected to remain finite in a regular horizon region.

Expanding our transverse matter work density in an orthonormal frame adapted to the privileged observers yields the algebraic identity:
\begin{equation}\label{omegaU}
\omega_{m}=\langle T_{\mu\nu}\rangle l^{+\mu}l^{-\nu}=\frac{1}{2}(\rho-p_{r})=p_{\perp}-\frac{1}{2}\langle{T^{\mu}}_{\mu}\rangle,
\end{equation}
where the last step is a straightforward algebraic identity derived from $\langle{T^{\mu}}_{\mu}\rangle=-\rho+p_{r}+2p_{\perp}$. Because $p_{\perp}$ and the trace are geometrically regular, the combination defining $\omega_m$ is expected to remain finite in a regular horizon limit. Therefore, the transverse variation operator $\delta$ intrinsically filters out the singular longitudinal fluxes driving the temporal evaporation so that the transverse matter work density remains well defined.

To explicitly evaluate this finite work density, one can consider the standard adiabatic \emph{approximation} for a macroscopic black hole. In this regime, the evaporation proceeds sufficiently slowly that the late-time, near-horizon physics is excellently modeled by the exact Unruh vacuum state ($|U\rangle$) evaluated on a fixed Schwarzschild background. 

The thermodynamic treatment of an actively evaporating black hole modeled by the Unruh state ($|U\rangle$) is traditionally obstructed by singularities in $\langle T_{\hat{\mu}\hat{\nu}}\rangle_{U}$ in the static frame when evaluated at the horizon. In effect, as explicitly demonstrated by Visser \cite{VisserU}, the static energy density $\rho$ and radial pressure $p_{r}$ measured by the privileged observers both diverge at the MTS due to the infinite relative boost of the observers in the limit as they approach the MTS. However, as shown above, our quasi-local framework algebraically bypasses this pathology.

In fact, in a fixed-background Schwarzschild approximation and when considering the action of conformally coupled massless scalars, the finite value of $\omega_{m}$ can be computed using the results in \cite{VisserU} (see appendix) as
\begin{equation}
\omega_{m}^{approx.}\simeq -42.25p_{\infty}.
\end{equation}
However, let us remark that the value obtained above belongs to the fixed-background Schwarzschild Unruh model and should therefore be viewed only as a leading-order input for an adiabatically backreacted semiclassical geometry.

Therefore, in this case one has $\delta W_{-}=\omega_{m}\delta V$, with a positive value (expected to be larger in magnitude than in the equilibrium case) and the symmetry again imposes $\delta Q_{-}=0$. Consequently, the transverse first law in a self-consistent backreacted geometry, free from the mathematical pathologies of temporal evaporative evolution, dictates that stepping transversally inward leaves a shell of this negative-energy quantum gas outside the control surface, triggering an increase in the quasi-local energy:
\begin{equation}
\delta E=\omega_{m}\delta V > 0.
\end{equation}

\section{Example: Non-spherically symmetric MTS}\label{secKerr}

Let us now analyze a non-spherically symmetric case: The case of a stationary rotating black hole. This is an interesting case to show the virtues of our approach since previous approaches \cite{Hay2004} needed a not yet found dual-null foliation of the spacetime which is now not required. 

Treating the general stationary rotating black hole is computationally demanding and will be treated in a future work. For this reason, in this section we will analyze the usual semiclassical fixed-background approximation in which the spacetime geometry is treated as Kerr at leading order. The nonzero effective temperature associated with the MTS motivates a semiclassical quantum state for matter fields on this background, with renormalized stress tensor $\langle T_{\mu\nu} \rangle = O(\hbar)$
Accordingly, the matter work density $\omega_m = \langle T_{\mu\nu} \rangle\, l_+^\mu l_-^\nu$
is generically nonzero, while the geometric quantities \(\kappa\), \(\omega_g\), \(E\), \(\Psi\), and \(\delta V\) will be evaluated using the Kerr background.

The Kerr metric can be written in advanced Eddington-Finkelstein-like coordinates as:
\begin{eqnarray*}
ds^2&=&-\left(1-\frac{2 M r}{\rho^2}\right) du^2+ 2\ du\ dr+ \rho^2 d\vartheta^2-\frac{4 a M r \sin^2\vartheta}{\rho^2} d\varphi\ du\\
& &-2 a \sin^2\vartheta\ d\varphi\ dr + \frac{(r^2+a^2)^2-a^2 \Delta \sin^2\vartheta}{\rho^2}\sin^2\vartheta d\varphi^2,
\end{eqnarray*}
where $\rho^2\equiv r^2+a^2 \cos^2\vartheta$ and
$\Delta\equiv r^2-2 r M+a^2$.

We choose to work with the surfaces $\mathcal S$ given by constant values of the coordinates $u$ and $r$.
Two obvious normal one-forms are given by $du$ and $dr$, and then the null normals, normalized with $l^{+}_{\mu}l^{-\mu}=-1$, can be given by
$$
\mathbf{\underline{l}}^+ = \frac{\rho^2}{2 \Theta^2} \left(-\Delta du +(r^2+a^2+ \Theta) dr \right) , \hspace{1cm} \mathbf{\underline{l}}^- = -du +\frac{r^2+a^2- \Theta}{\Delta} dr
$$
where $\Theta$ is a shorthand for
$$
\Theta = \sqrt{(r^2+a^2)\rho^2+2Mra^2\sin^2\theta} .
$$
Observe that at the roots of $\Delta =0$ one also has $r^{2}+a^{2}-\Theta =0$ and thus the $dr$-component of $\mathbf{\underline{l}}^-$ can be well defined even if those roots exist (what happens whenever $M^2 \geq a^2$).
The expansion corresponding to the outgoing null geodesics is
\begin{equation}
\theta^+ =\frac{\Delta}{2 \Theta^3}  [2r\rho^2+a^2\sin^2\theta (M+r)]
\end{equation}
From this expression, we deduce that the chosen surfaces $\mathcal S$ are marginally trapped when $\Delta=0$. In other words, as is well-known, provided that $M^{2}>a^2$, for every value of the coordinate $u$ there are two MTS defined by the values of the coordinate $r$
\begin{equation}\label{Horizon}
r_{\pm}=M\pm \sqrt{M^2-a^2}.
\end{equation}

Since the surfaces defined by constant values of $u$ and $r$ are compact, we can write the effective surface gravity associated with these MTSs as
\begin{eqnarray*}
\kappa&=& -\sqrt{\frac{A(\mathcal S_{+})}{16\pi}}\  l^{-\alpha}\nabla_\alpha \theta^+\rfloor_{+}=\\
&=&\frac{(r_+ - M) [4 r_+^2 \rho_+^2+a^2(a^2+3 r_+^2)\sin^2\theta]}{4 r_+ (a^2+r_+^2)^{3/2} \rho_+^2},
\end{eqnarray*}
where the subscript `$+$' indicates that the quantity is evaluated on the MTS $(u_0, r_+)$ and we have used that the area of the MTS is $A(\mathcal S_{+})=4\pi (r_{+}^2+a^2)$. The mean effective surface gravity on the MTS can now be computed as
\begin{eqnarray*}
\overline\kappa &=&\frac{1}{A} \oint_{\mathcal S} \ast\ \kappa=\\
&=& \frac{(r_+-m) [a r_+^3-a^3 r_+ + (a^2+r_+^2)(a^2+3 r_+^2) \tan^{-1}(a/r_+)]}{4 a r_+^2 (a^2+r_+^2)^{3/2}}.
\end{eqnarray*}

By using (\ref{zeta}), we get on $\mathcal S_+$
\[
|\zeta|^2=\frac{a^2 \sin^2\vartheta [a^6+(a^3-a r_+^2)^2 \cos(2\vartheta)+13 a^2 r_+^4+18 r_+^6]}{2 r_+^2 (a^2+r_+^2)^2 (a^2 \cos(2 \vartheta)+a^2+2 r_+^2)^2}.
\]
which, with the help of (\ref{omg}) and (\ref{mean}), allows us to find the mean gravitational work density on the MTS as
\[
\overline\omega_{g}=\frac{3 a r_+^5-a^5 r_++6 a^3 r_+^3+(a-r_+)(a+r_+)(a^2+r_+^2)(a^2+3 r_+^2) \tan^{-1}(a/r_+)}{32 \pi a r_+^3 (a^2+r_+^2)^2}.
\]

Thus, unlike in the previously treated spherically symmetric case, these surfaces have nonzero twist $\zeta\neq 0$ and, consequently, nonzero gravitational work densities: $\omega_g \neq 0$ and $\overline\omega_g \neq 0$.

The dual expansion vector
takes the explicit form
\[
{}^*H
=
\frac{N}{\rho^2\Theta}
\left(
\partial_u+\frac{2aMr}{\Theta^2}\,\partial_\varphi
\right),
\]
where $N \equiv 2r\rho^2+a^2(M+r)\sin^2\vartheta$.

Therefore, 
in this case, the 4-velocity of the privileged observers is
\[
\hat u
=
\frac{\Theta}{\rho\sqrt{\Delta}}
\left(
\partial_u+\frac{2aMr}{\Theta^2}\,\partial_\varphi
\right).
\]

Note that the privileged observers are locally rotating and their angular velocity is
\[
\Omega_{\rm pr}=\frac{2aMr}{\Theta^2},
\]
which in general depends on $r$ and $\theta$. 
On the MTS $(r=r_+,\Delta=0)$, however, one has $\Theta^2=(r_+^2+a^2)^2$ and $2Mr_+=r_+^2+a^2$, so that
\[
\Omega_{\rm pr}\big|_{r_+}=\frac{a}{r_+^2+a^2}=\Omega_H=\text{constant},
\]
where $\Omega_H$ is the angular velocity of rigidly corotating observers.
Hence both observer fields become asymptotically parallel to the same null horizon generator $\chi=\partial_u+\Omega_H\partial_\varphi$ as the MTS is approached. Clearly, the Lorentz scalar $\omega_m=T_{\mu\nu}l_+^\mu l_-^\nu$ has the same MTS value whether it is rewritten in the limiting privileged frame or in the limiting corotating frame.
Moreover, because the kinematic frames of the privileged and corotating observers asymptotically converge at the MTS, the physical components of the stress-energy tensor measured by both observers at the MTS (such as the local energy density $\rho$) 
agree in the limit as $r\rightarrow r_+$ from the exterior.
This geometric alignment rigorously ensures that extracting physical components from the rigidly corotating thermal cavity remains flawlessly consistent with our underlying quasi-local framework.

\subsection{Kerr black hole in semiclassical equilibrium}

To achieve strict thermal equilibrium, the Kerr black hole must be enclosed within a perfectly reflecting cavity maintained at the co-rotating angular velocity $\Omega_H = a/(r_+^2+a^2)$. While for a spherically symmetric black hole a cavity is merely a thermodynamic device utilized to stabilize the canonical ensemble against a negative heat capacity, for a rotating Kerr black hole, the bounding box is a strict geometric and quantum field-theoretic necessity. 

As is rigorously established by the Kay-Wald theorem \cite{KayWald1991}, a globally regular ``thermal, rigidly rotating'' state cannot exist on the full unbounded exterior spacetime. The horizon-generating Killing vector field $\chi=\partial_u+\Omega_H\partial_\varphi$ naturally becomes spacelike at large radii, forming a ``speed-of-light surface'' where a co-rotating thermal plasma would be forced to exceed the speed of light. Therefore, a true macroscopic equilibrium can only be achieved by imposing a reflecting boundary on a domain strictly before this superluminal surface. Confining the system within such a subluminal cavity alters the boundary-value problem and allows for the rigorous construction of a regular, Hartle-Hawking-like equilibrium state \cite{DuffyOttewill2008}. 

In the corotating local Lorentz frame, regularity on the (non-extremal) horizon implies the
same kinematic constraint in the normal 
plane as in subsec.\ref{subsecEquil}:
\begin{equation}
p_r = -\rho,
\end{equation}
so that the matter work density on the MTS is
\begin{equation}\label{omeq}
\omega_m \equiv \langle T_{\mu\nu}\rangle_{\rm eq}\,l_+^\mu l_-^\nu
= \frac{1}{2}(\rho-p_r)=\rho.
\end{equation}

Let us remark that this equation is to be understood on the MTS, or equivalently as the limit $r\to r_+$ from the exterior.

In this case, the local matter work density $\omega_m(\vartheta) = \rho(\vartheta)$ is distributed unevenly from the poles to the equator. Explicit numerical evaluations of this latitude-dependent profile for a Hartle-Hawking-like cavity state can be found in \cite{DuffyOttewill2008}.
It is precisely here that the integral formulation of our transverse first law demonstrates its utility. 
By integrating the local matter work density $\omega_m(\vartheta)$ alongside the non-uniform gravitational work density $\omega_g(\vartheta)$ across the closed 2-surface, the mean work density $\bar{\omega}$ rigorously captures the net macroscopic energetic cost of the deformation, bypassing the need to homogenize the fundamentally asymmetric quantum plasma.

Meanwhile the transverse variation of generalized heat (\ref{dQ-k}) 
satisfies $\delta Q_- \neq 0$. Therefore, according to the transverse first law (\ref{firstlaw}), the variation of energy as one traverses the MTS along the tube receives both contributions $\delta E\stackrel{MTS}{=} \delta Q_- + \delta W_-$.

\subsection{Evaporating Kerr black hole}

In the Unruh state, the non-trivial angular-flux components $(\langle T_{u\varphi}\rangle_U,\langle T_{r\varphi}\rangle_U)$ are nonzero. A direct evaluation of the matter work density $\omega_m=\langle T_{\mu\nu}\rangle_U l^{+\mu}l^{-\nu}$
therefore appears, a priori, to involve these angular-momentum fluxes, since the null normal $l^{-\mu}$ has a nonvanishing azimuthal component $(l^{-\varphi}\neq0)$. This difficulty can be avoided algebraically by using the exact decomposition of the inverse metric in terms of the null normals and the induced 2-metric $\gamma^{\alpha\beta}$:
\[
g^{\alpha\beta}=-l^{+\alpha}l^{-\beta}-l^{-\alpha}l^{+\beta}+\gamma^{\alpha\beta}.
\]

Taking the four-dimensional trace of the renormalized stress-energy tensor then gives
$\langle T^\mu{}_\mu\rangle_U=-2\omega_m+\gamma^{\alpha\beta}\langle T_{\alpha\beta}\rangle_U$.
Defining the intrinsic transverse pressure trace by $P_\perp\equiv \frac12\,\gamma^{\alpha\beta}\langle T_{\alpha\beta}\rangle_U,$
one obtains (generalizing (\ref{omegaU})
\begin{equation}\label{eq:bypass_kerr}
\omega_m=P_\perp-\frac12\langle T^\mu{}_\mu\rangle_U.
\end{equation}
Equation (\ref{eq:bypass_kerr}) therefore expresses $\omega_m$ in terms of the intrinsic transverse pressure trace $P_\perp$ and the scalar trace $\langle T^\mu{}_\mu\rangle_U$, without requiring an explicit evaluation of the longitudinal angular-flux sector.

Recent landmark numerical computations 
\cite{LeviEilonOriVandeMeent2017} have successfully evaluated the full renormalized stress-energy tensor for the Unruh state at the event horizon (specifically for a spin of $a=0.7M$). Their numerical results, obtained for the specific Unruh-state configuration under consideration, indicate that the transverse components of the RSET remain finite on the horizon at all latitudes.
In our framework, feeding their numerically generated, latitude-dependent pressure trace $\mathcal{P}_\perp(\vartheta)$ directly into our algebraic identity (\ref{eq:bypass_kerr}) would yield the exact local profile of the matter work density $\omega_m(\vartheta)$ point-by-point across the MTS. 
As in the equilibrium case, the variation of total work receives contributions both from the matter work density and the gravitational work density: $\delta W_-=(\bar \omega_m+ \bar \omega_g) \delta V$. 
The variation of generalized heat $\delta Q_-$ takes the same non-zero value as in the equilibrium case within the unchanged Kerr geometry approximation. Therefore, according to the transverse first law (\ref{firstlaw}), the variation of energy as one traverses the MTS along the tube receives both contributions $\delta E\stackrel{MTS}{=} \delta Q_- + \delta W_-$. 
The transverse first law thus seamlessly incorporates state-of-the-art numerical QFT data.
The framework isolates a transverse sector in which the relevant matter contribution remains well defined even in the evaporating regime.

\section{Conclusions}\label{secConc}

In this work we have formulated an analogue transverse first law for a general closed marginally trapped surface (MTS), attached directly to the surface itself rather than to a preferred horizon worldtube. Taking the Hawking energy $E$ as the quasi-local internal energy and the effective surface gravity $\kappa$ as the quantity controlling the thermal sector, we obtained an exact deformation law in which the ingoing-null variation of $E$ splits into a generalized heat term $\delta Q_-$ and a total work term $\delta W_-$:
\[
\delta E\stackrel{MTS}{=} \delta Q_- + \delta W_-.
\]
In this way, the formalism provides a genuinely codimension-two and quasi-local counterpart to horizon-based laws, while remaining applicable even when no unique preferred horizon hypersurface is available or when several horizon segments coexist. This is precisely the setting in which an individual MTS, rather than a selected three-dimensional horizon tube, becomes the natural carrier of the relevant thermodynamic data.

The physical content of the two terms is clear. The work contribution $\delta W_-=\bar\omega\,\delta V$, with $\bar\omega=\bar\omega_m+\bar\omega_g$, measures the total matter-plus-geometric work associated with the transverse displacement of the control surface. In spherical symmetry the geometric contribution vanishes and one recovers purely matter work, whereas in Kerr the twist of the null normals makes $\bar\omega_g\neq 0$, showing that rotation introduces an intrinsically geometric work channel even before backreaction is included. The heat contribution is subtler. It is not exhausted by the familiar $(\bar\kappa/8\pi)\,\delta A$ term, because it also contains $R\,\delta\Psi$, which accounts for the variation of the quasi-local concentration of gravitational energy. Most revealingly, the transverse variation of heat assumes a standard macroscopic non-equilibrium thermodynamic form when expressed as an integral over the local geometric entropy variation
\[
\delta Q_-=\int_{\rm MTS}(\bar T-T)\,\delta s,
\]
so that $\delta Q_-$ is most cleanly interpreted as an entropy-weighted measure of transverse temperature inhomogeneity on the MTS. It vanishes for uniformly symmetric MTSs, for which the effective temperature is constant over the surface, and becomes nonzero when the surface is geometrically and thermodynamically distorted. 

It is also essential to state clearly the mathematical nature of the operator $\delta$. Here $\delta=l_-^\alpha\nabla_\alpha$ is not a phase-space variation comparing neighboring solutions, as in the covariant phase-space/Noether-charge framework or in isolated-horizon Hamiltonian mechanics \cite{IyerWald1994,AshtekarFairhurstKrishnan2000}. Rather, it is an exact deformation operator defined within a single spacetime, evaluating how the quasi-local quantities attached to a given MTS change under an inward null displacement. The resulting relation is, therefore, a local transverse deformation identity, not a mechanical first law in the strict phase-space sense. This distinction is not merely terminological: it is precisely what allows the balance law to remain meaningful without assuming global equilibrium.

This point also clarifies the relation of the present framework to recent literature on quasi-local horizons. Modern dynamical-horizon and far-from-equilibrium analyses provide longitudinal evolution laws along selected horizon segments and have greatly expanded the thermodynamics of nonstationary black holes, including evaporating timelike phases \cite{AshtekarKrishnan2025,AshtekarParaizoShu2025}. Our result does not replace those constructions; it complements them by isolating the transverse, instantaneous balance carried by each individual MTS. In this sense, the present formalism supplies the codimension-two counterpart to the longitudinal laws emphasized in recent quasi-local-horizon work.

The semiclassical examples studied here support that interpretation. For round spheres in spherically symmetric spacetimes, the formalism reproduces the expected adiabatic result $\delta Q_-=0$, both in equilibrium and in the evaporating case, while the work term correctly tracks the matter contribution. More importantly, in the evaporating regime the framework avoids the usual obstruction associated with the singular longitudinal Unruh fluxes in static frames. The reason is structural: the work density entering the law is $T_{\mu\nu}l_+^\mu l_-^\nu$, not the longitudinal flux $T_{\mu\nu}l_+^\mu l_+^\nu$ that drives the shrinking of the hole. In spherical symmetry this transverse contraction can be rewritten in terms of the transverse pressure and the scalar trace, yielding a horizon-finite quantity; the Kerr analysis exhibits the same basic decoupling from the longitudinal sector, even though the rotating case is technically more involved. Thus, the transverse law remains well defined as an instantaneous quasi-local balance even in regimes where longitudinal evolution is cumbersome to handle.

Several directions for further work follow naturally. A first and conceptually clean extension is to apply the formalism to non-round, non-spherically symmetric MTSs embedded in spherically symmetric spacetimes. This would isolate the role of the intrinsic distortion of the MTS from that of the ambient spacetime geometry, providing an intermediate step between the round-sphere sector and the genuinely nonspherical Kerr case. A second priority is a fully self-consistent treatment of rotating black holes including semiclassical backreaction, beyond the fixed-background Kerr approximation used here. A third direction concerns mode-by-mode analyses. Strictly speaking, what should be decomposed is not the already integrated scalars $\delta Q_-$ and $\delta W_-$ themselves, but the local scalar densities whose integrals define them. For topologically spherical MTSs, one may project those densities onto an intrinsic harmonic basis and study how the corresponding thermodynamic multipoles relax toward equilibrium. This would provide a natural bridge between the present framework and the dynamical-horizon multipole program developed in recent numerical-relativity studies \cite{AshtekarKrishnan2025,AshtekarCampigliaShah2013,ChenEtAl2022}.

Finally, the assignment of temperature and entropy to every MTS invites, but does not force, a microscopic reading. The present analysis is macroscopic and quasi-local: it shows that a general MTS supports a thermodynamic balance law, but it does not identify the underlying degrees of freedom. Still, this conclusion is compatible with the broader idea---suggested in Bekenstein-type quantization pictures, in loop-quantum-gravity isolated-horizon state counting, and in string-theoretic microstate constructions---that black-hole thermodynamics admits a statistical interpretation \cite{BekensteinMukhanov1995,AshtekarBaezCorichiKrasnov1998,StromingerVafa1996}. At present, the cautious statement is that the present framework supplies a quasi-local macroscopic framework on which any future microscopic account of black-hole degrees of freedom should be able to act.

Taken together, these results support the view that closed marginally trapped surfaces provide a natural arena for a genuinely quasi-local thermodynamics of black holes: one that is exact at the level of transverse deformation identities, flexible enough to accommodate symmetry breaking and rotation, and structurally robust in both semiclassical equilibrium and evaporation.

\section*{Appendix}

To explicitly prove how the operator $\delta$ bypasses the divergences at the horizon in the fixed-background Schwarzschild approximation for the evaporating case, we utilize Visser's exact covariant decomposition of the Unruh stress-energy tensor \cite{VisserU}. Defining $z \equiv 2 m/r$, Visser proves that the exact functional forms of the longitudinal components are:
\begin{align}
\rho(z) &= -f_0 \frac{z^2}{1-z} + \frac{\xi}{10} p_\infty z^2 (1+z+z^2+z^3-9z^4) + p_\perp(z) - \frac{z^2}{1-z}F(z) \\
p_r(z) &= -f_0 \frac{z^2}{1-z} + \frac{\xi}{10} p_\infty z^2 (1+z+z^2+z^3+z^4) - p_\perp(z) - \frac{z^2}{1-z}F(z)
\end{align}
where $f_0$ governs the outgoing singular Hawking flux, $F(z)$ is a finite non-local pressure integral, $\xi$ is the trace anomaly coefficient (e.g., $\xi=96$ for a scalar field), and $p_\infty$ is the asymptotic pressure scale. The explicit pole $(1-z)^{-1}$ drives the divergence at the MTS ($z \to 1$). 

Crucially, when we compute our transverse matter work density $\omega_m = \frac{1}{2}(\rho - p_r)$, the divergent flux terms and the non-local integrals mathematically annihilate each other identically point-by-point:
\begin{equation}
\omega_m = 
p_\perp(z) - \frac{1}{2} \xi p_\infty z^6.
\end{equation}
Recognizing that the geometric conformal trace anomaly is exactly $\langle T^\mu_\mu \rangle_U = \xi p_\infty z^6$, we explicitly recover our globally regular algebraic identity.  

Using Visser's semi-analytic model fitted to numerical QFT data \cite{VisserU}, we can explicitly evaluate this work density for a conformally coupled massless scalar field at the MTS ($z=1$). With $\langle T^\mu_\mu \rangle_U = 96 p_\infty$ and $p_\perp \approx 5.75 p_\infty$, the matter work density on the MTS takes the finite, negative value:
\begin{equation}
\omega_m\rfloor_{MTS} \approx -42.25 p_\infty.
\end{equation}

Therefore, the transverse variation operator $\delta$
removes any explicit dependence on the longitudinal mass-loss fluxes and on the angular-flux sector associated with superradiant scattering.

\end{document}